\documentclass[twocolumn,tighten,times]{aastex62}
\usepackage{xspace}
\usepackage{xcolor}
\usepackage{graphicx}
\usepackage[normalem]{ulem} 
\newcommand{\ampl}{x}
\newcommand{\dv}{\!\:{/}\!\:} 
\newcommand{\eq}{\!\:{=}\!\:} 
\newcommand{\pls}{\!\:{+}\!\:} 
\newcommand{\mns}{\!\:{-}\!\:} 
\newcommand{\lt}{\!\:{<}\!\:} 
	
\newcommand{\mycite}[1]{{\color{blue}\cite{#1}}}
\newcommand{\mycitep}[1]{{\color{blue}\citep{#1}}}
\newcommand{\mycitet}[1]{{\color{blue}\citet{#1}}}
\newcommand{\eqref}[1]{{\color{blue}Equation\,\ref{#1}}} 
\newcommand{\figref}[1]{{\color{blue}Figure\,\ref{#1}}}
\newcommand{\tabref}[1]{{\color{blue}Table\,\ref{#1}}}
\newcommand{\appref}{{\color{blue}Appendix}\xspace}
\newcommand{\ud}[1]{\mathrm{d}{#1}} 
\newcommand{\br}[1]{\left(#1\right)} 
\newcommand{\xiplus}{(1{+}\ampl)} 
\newcommand{\km}{\mathrm{km}}
\newcommand{\eV}{\mathrm{eV}}
\newcommand{\keV}{\mathrm{keV}}
\newcommand{\MeV}{\mathrm{MeV}}
\renewcommand{\fm}{\mathrm{fm}}
\newcommand{\icfm}{\mathrm{fm}^{\!{-}\!3}}
\newcommand{\icc}{\mathrm{cm}^{\!{-}\!3}}
\newcommand{\gcc}{\mathrm{g}{\cdot}\mathrm{cm}^{\!{-}\!3}}
\renewcommand{\sec}{\mathrm{s}}
\newcommand{\kmisec}{\km{\cdot}\sec^{\!{-}1}}
\newcommand{\erg}{\mathrm{erg}}
\newcommand{\ergisec}{\erg{\cdot}\sec^{\!{-}1}}
\newcommand{\ergisecicc}{\erg{\cdot}\sec^{\!{-}\!1}\!{\cdot}\icc}
\newcommand{\ergisecifm}{\erg{\cdot}\sec^{\!{-}\!1}\!{\cdot}\mathrm{fm}^{\!{-}\!1}}
\newcommand{\mevicfm}{\mathrm{MeV}{\!\!\;\cdot}\mathrm{fm}^{\!{-}\!3}}
\newcommand{\sqm}{{SQM}\xspace}
\newcommand{\msun}{{M}_{\odot}}
\newcommand{\mearth}{{M}_{\oplus}}

\newcommand{\isec}{\mathrm{s}^{\!{-}\!1}}
\newcommand{\msek}{\mathrm{m{}s}}
\newcommand{\mb}{\mathrm{mb}}
\newcommand{\nb}{N_{\!\!\:\;\!\!{B}}}
\newcommand{\gammg}{{\Gamma}_{\!\!\:\;\!\!{G}}}
\newcommand{\cm}{\mathrm{cm}}
\newcommand{\m}{\mathrm{m}}
\newcommand{\ms}{m_{\mathrm{s}}}

\newcommand{\Teff}{T_\mathrm{eff}}

\newcommand{\Emin}{\mathrm{E}_\mathrm{min}}
\definecolor{mygreen}{RGB}{15,135,30}
\newcommand{\erad}{E_\mathrm{T}}

\definecolor{myorange}{RGB}{180,0,220}
\newcommand{\epsrate}{P}
\newcommand{\epsq}{\varepsilon_{\!\!\;q}} 
\newcommand{\epsnu}{\varepsilon_{\!\!\;\nu}} 
\newcommand{\epsat}{\epsilon_{\!\!\;s}} 
\newcommand{\excen}{{\epsilon}_{\mathrm{exc}}}
\newcommand{\meanexcen}{\bar{\epsilon}_{\mathrm{exc}}}

\definecolor{newcitecolor}{RGB}{255,0,255}

\newcommand{\dR}{\delta\!R_{\ampl}}

\begin{document}
\title{{\bf\large Oscillating Strange Quark Matter Objects Excited in Stellar Systems}}
\author{Marek Kutschera}
\affiliation{{\scriptsize Jagiellonian University, {\L}ojasiewicza 11,  PL-30348  Krak\'{o}w, Poland}}
\author{Joanna Ja{\l}ocha}
\affiliation{{\scriptsize Institute of Physics, Faculty of Materials Engineering and Physics, Cracow University of Technology, Podchor\c{a}\.{z}ych 1, PL-30084 Krak\'{o}w, Poland}}
\author{{\L}ukasz Bratek}
\affiliation{{\scriptsize Institute of Physics, Faculty of Materials Engineering and Physics, Cracow University of Technology, Podchor\c{a}\.{z}ych 1, PL-30084 Krak\'{o}w, Poland}}
\author{Sebastian Kubis}
\affiliation{{\scriptsize Institute of Physics, Faculty of Materials Engineering and Physics, Cracow University of Technology, Podchor\c{a}\.{z}ych 1, PL-30084 Krak\'{o}w, Poland}}
\author{Tomasz K\c{e}dziorek}
\affiliation{{\scriptsize Jagiellonian University, {\L}ojasiewicza 11,  PL-30348  Krak\'{o}w, Poland}}
\correspondingauthor{{\L}ukasz Bratek}
\email{lukasz.bratek@pk.edu.pl}

\begin{abstract}
It is shown that strange quark matter (\sqm) objects, stars, and planets, can very efficiently convert the mechanical energy into hadronic energy when they oscillate. 
This is because the mass density at the edge of \sqm objects,  $\rho_0\eq4.7{\times}10^{14}\,\gcc$, is the critical density below which \sqm is unstable with respect to decay into photons, hadrons, and leptons. 
We consider here radial oscillations of \sqm objects that could be induced in stellar or planetary systems where tidal interactions are ubiquitous. 
Oscillations of $0.1\%$ radius amplitude already result in $1\,\keV$ per unit baryon number excitation near the surface of \sqm stars. 
The excitation energy is converted into electromagnetic energy in a short time of $1\,\msek$, during a few oscillations. 
Higher amplitude oscillations  result in faster energy release that could lead to fragmentation or  dissolution of \sqm stars. 
This would have significant consequences for hypothetical \sqm star binaries and planetary systems of \sqm planets with regard to gravitational wave emission.
\end{abstract}


\section{Introduction}
Recent papers \mycitep{2017ApJ...848..115H,Kuerban_2020} renew the interest in strange quark matter (\sqm) objects of planetary masses \mycitep{2010arXiv1010.2056K} as candidates for high density planets orbiting some pulsars. 
Also, \sqm planets around neutron stars are proposed to be new sources of gravitational radiation that could be detected by new generation detectors \mycitep{2015ApJ...804...21G}.
	
The idea that \sqm forms the ground state of baryon matter was put forward by \mycitet{1984PhRvD..30..272W}. 
Neutron stars as \sqm stars were studied soon after \mycitep{1986ApJ...310..261A}. 
It was pointed out that there could exist \sqm objects of any mass down to planetary values \mycitep{1996astro.ph..4035W} and even smaller objects, called strange nuggets \mycitep{1984PhRvD..30.2379F}, could be present in the space.
	
In \mycitet{2017ApJ...848..115H} and \mycitet{Kuerban_2020} the pulsar PSR B1257+12 is proposed as one of the hosts of candidates for \sqm planets. 
It is the first stellar object discovered to harbor three planets \mycitep{1992Natur.355..145W} of $4.3\,\mearth$, $3.9\,\mearth$, and $0.025\,\mearth$. 
This system was conjectured in \mycitet{2010arXiv1010.2056K} to be composed of \sqm. 

Here our aim is to study the dynamical behavior of \sqm stars and planets interacting with other objects in stellar systems, in particular, stability of \sqm objects with respect to radial oscillations. 
Such oscillations occur when these objects are subject to tidal interactions in stellar or planetary systems, especially during the formation of binary systems and close encounters with other stars. 
Here we assume that the oscillations of an \sqm  star are excited during the close encounter with another neutron star or a black hole \mycitep{2018PhRvD..98d4007Y}. 

Radial oscillations of compact stars (i.e. neutron stars and quark stars) have
been studied in \mycitet{1983ApJS...53...93G} and \mycitet{1990ApJ...363..603C} (see also \mycitet{1992A&A...260..250V,2001A&A...366..565K}). 
For a number of equations of state of dense matter the fundamental mode frequencies are listed as functions of stellar mass \mycitep{1983ApJS...53...93G,1990ApJ...363..603C}. 
For the \sqm star of $1.4\,\msun$ the fundamental mode frequency is calculated  for the \sqm equation of state due to  \mycite{1989PhRvL..63.2629G} to be $\omega\eq16900\,\isec$ and the period of oscillations is $T\eq0.37\,\msek$ \mycitep{1990ApJ...363..603C}. 
Periods of a few tenths of millisecond are typical  for  other equations of state. 
The energy of such oscillations can be estimated to be $~(\xi{/}R)^2{\times}10^{53}\,\erg$, where $R$ is the stellar radius and $\xi$ is the amplitude of surface displacement \mycitep{1983ApJS...53...93G}. 
	
The \sqm stars and planets are very compact objects with a radius of $10.3\,\km$ for a star of $1.4\,\msun$ to $145\,\m$ for the planet of $1\,\mearth$. 
\sqm stars have much more uniform distribution of mass as compared to normal neutron stars. 
The central densities for both branches of stars are similar, of order $10^{15}\,\gcc$. 
However, neutron stars have low density crust and the surface (defined by the condition the pressure vanishes there) is of density of a few $\gcc$ \mycitep{1971ApJ...170..299B}. 
In contrast, the surface of \sqm stars is of density  $\rho_0\eq4.665{\times}10^{14}\,\gcc$. 
Of course, the same is for \sqm planets (we consider here bare \sqm stars for simplicity with no crust, which could be present \mycitep{1986ApJ...310..261A}). 
The density  $\rho_0$  is the saturation density of strange quark  matter \mycitep{1984PhRvD..30.2379F} and it is the crucial parameter for \sqm objects.
	
	\section{PROPERTIES OF \sqm STARS AND PLANETS}

	The properties of  \sqm are calculated here in a simple model approach within the MIT-bag scenario \mycitep{2010arXiv1010.2056K} and are parameterized mainly by the value of the bag constant $B$.
The energy density, $\rho\,c^2\eq\,\epsq{+}B$, 
includes relativistic energy density of quarks $\epsq$ defined by the requirement that chemical potentials of electrically neutral matter  satisfy the beta-stability condition. 

\begin{figure}[h]
\includegraphics[trim=0 0 0 0,clip,width=0.95\columnwidth]{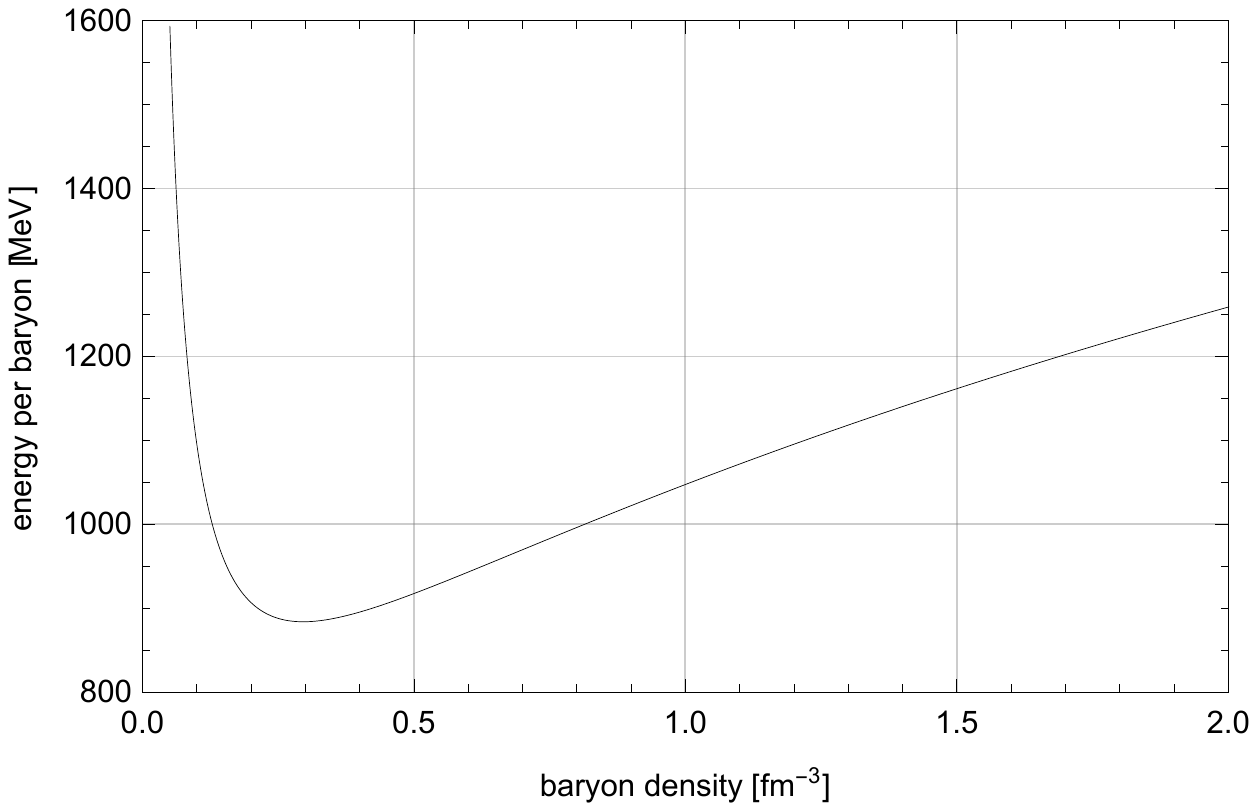}%
\caption{\label{fig:energyperbaryon}\small Energy per baryon versus baryon density for \sqm \\ ($B=60\,\mevicfm$, $\ms=150\,\MeV$).}
\end{figure}

In what follows we assume $B\eq60\,\mevicfm$, and strange quark mass $m_s\eq150\,\MeV$ (with {\it up/down} quarks and electrons being massless). 
The resulting energy per baryon is shown in \figref{fig:energyperbaryon}.	
In our model the saturation density (at the minimum) is $\rho_0\eq4.665{\times}10^{14}\,\gcc$. 

The saturation density  $\rho_0$ is the critical density for \sqm because baryon matter is stable only as long as its density $\rho {>}\rho_0$. 
At lower densities, $\rho{{{<}}}\rho_0$, baryon matter decays into leptons, photons, and hadrons conserving the baryon number. 
The baryon density at the saturation is $n_0\eq0.2961\,\icfm$ and the energy per baryon  has its minimum $\Emin\eq883.6\,\MeV$. 
For lower baryon densities, $n{{<}}n_0$, the energy per baryon increases.
		
\begin{figure}
\includegraphics[trim=0 19 0 0,clip,width=0.93\columnwidth]{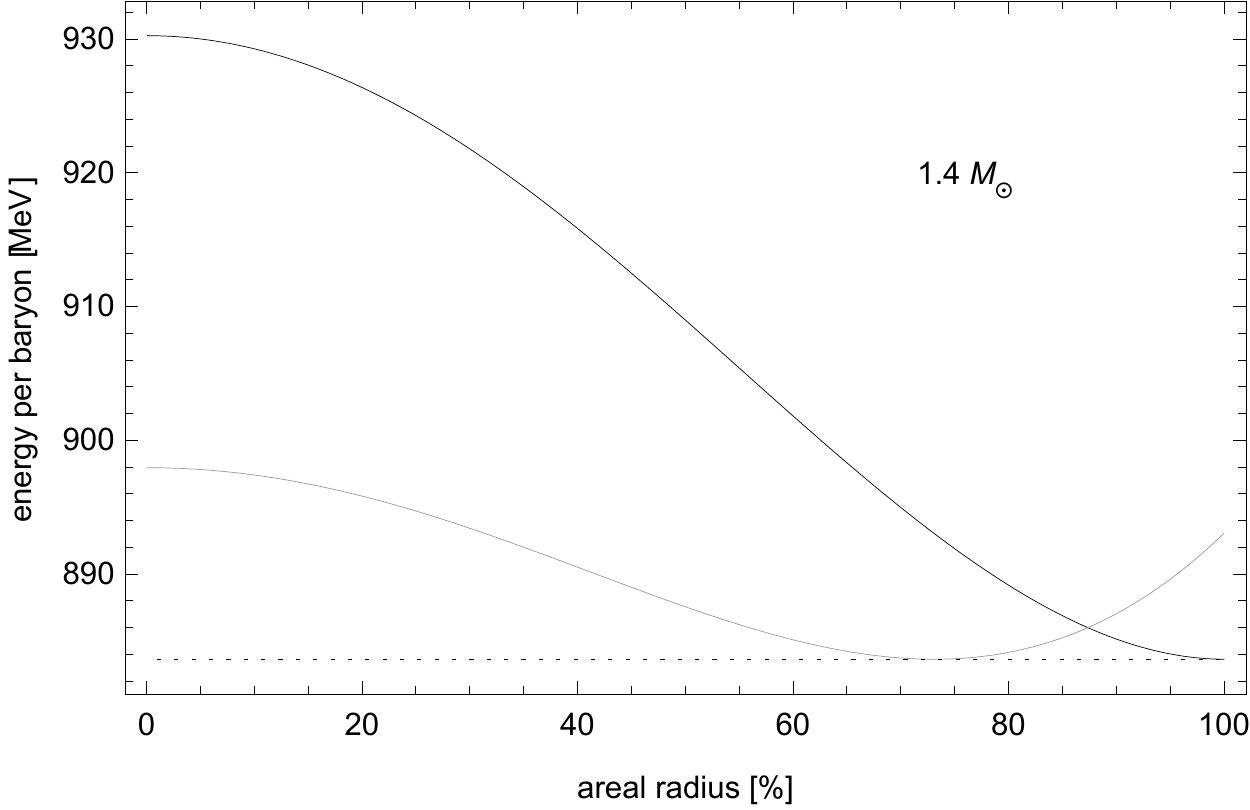}\\
\includegraphics[trim=2 0 0 0,clip,width=0.93\columnwidth]{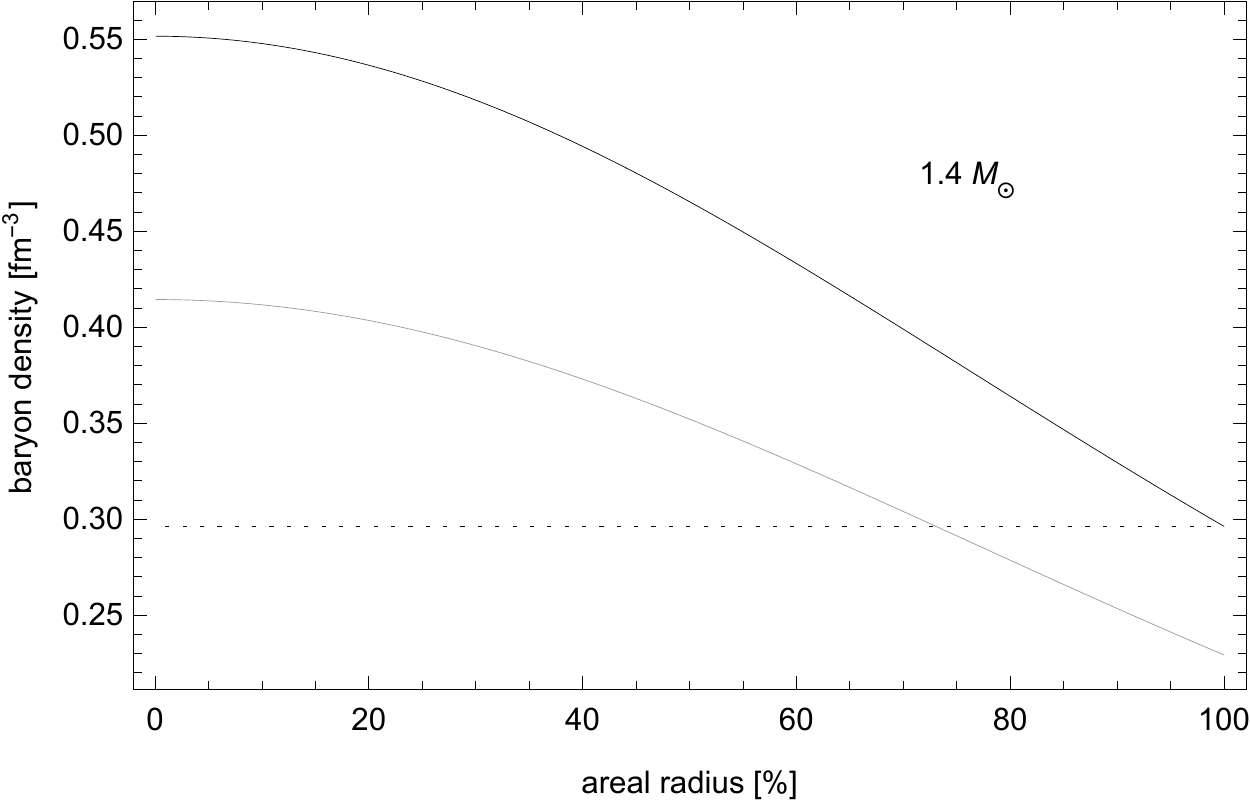}
\caption{\label{fig:star}\small Baryon density and energy per baryon distribution in \sqm generally-relativistic star of mass $1.4\msun$. 
Both panels: {\it black lines} -- the equilibrium state solution, {\it gray lines} -- radially scaled solution at $10\%$ amplitude of radial oscillations, {\it dashed lines} -- saturation baryon density and the corresponding energy per baryon. }
\end{figure}

In \figref{fig:star} we show distribution of the baryon density in the \sqm generally-relativistic spherical star of mass $1.4\,\msun$. 
The areal radius of the star is $R\eq10.31\,\km$. 
Also the scaled distribution corresponding to the maximum expansion of the star undergoing $10\%$ radial oscillations is shown. 
The oscillation amplitude  of $10\%$ (which is quite large) is chosen for the sake of illustration. 
In the following we focus on smaller amplitudes. 
In the scaled distribution, which is meant to model a uniform expansion of the oscillating star, the baryon density drops below the saturation value $n_0$ in the outer shell of the star with radii $R{>}r{>}r_c$ (here, $r_c\eq7.535\,\km$). 
In the top panel in \figref{fig:star} one can  see the energy per baryon as a function of areal radius. 
The minimum energy per baryon occurs at $r\eq r_c$ and then the energy increases toward the surface where the value higher by $9.450\,\MeV$ is found.
	
One can imagine how vulnerable the surface  of \sqm objects is to  even small fluctuations. 
The energy per unit baryon number at the surface  for $0.1\%$ amplitude radial oscillation is calculated to be $1.221\,\keV$ above the minimum energy $\Emin$. 
This corresponds to radius expansion to $R{+}\xi$, where $\xi\eq10.31\,\m$ for $R\eq10.31\,\km$, or $\ampl\eq\xi/R\eq0.001$. 
The baryon density decreases to roughly $n_0/(1{+}\ampl)^3\eq0.2953\,\icfm$ at the surface. 
The shell of $n{<}n_0$ comprises the outer $26\,\m$ of the star's areal radius.
	 
	 Generally, \sqm becomes excited as baryon density is lowered below saturation density $n_0$ and the excess energy could finally be radiated away. 
We wish to stress here that the excitation effect occurs only in radial oscillations of \sqm stars. 
It is absent for normal neutron stars. 
For the sake of argument assume that the crust of a neutron star is an iron crystal \mycitep{1971ApJ...170..299B} of baryon density $n\eq4.73{\times}10^{24}\,\icc$. 
The change to $n/(1{+}\ampl)^3\eq4.716{\times}10^{24}\,\icc$ with $\ampl\eq0.001$ hardly makes any difference to the system: the iron crystal becomes slightly less compressed.

There are situations when the surface of an \sqm object can be stationary. 
This could happen for single, isolated stars or planets that are cold. 
In stellar and planetary systems, where astrophysical objects interact with one another, a fully static surface is difficult to imagine. 
Here we consider what can happen when the surface of an \sqm star oscillates radially (in so-called monopole or breathing mode).

The radial oscillations engulf the whole star. 
All elements of mass oscillate in phase with the same frequency around equilibrium positions. 
At the maximum expansion the radius of the star increases to $R{+}\xi$ and at the maximum compression it decreases to $R{-}\xi$, where $R$ is the equilibrium radius of the star. 
The volume of the star oscillates and so does the density. 
At the maximum expansion the density is lowest. 
                     
\begin{table*}
\centering
  \begin{small}
\begin{tabular}{|c|c|c|c|c|c|c|c|}
\hline
Mass & $\nb$ & $R$ & $\ampl$ & $\erad$ & $r_c$ & $\nb(r{>}r_c)$ & $\meanexcen \left.(\excen\right|_R)$ \\
\hline
\hline
$1.0\,\mearth$ & $3.790{\times}10^{51}$ & $145.1\,\m$ & $0.001$ & $7.315{\times}10^{42}\,\erg$ & $0.0\,\m$ &  $3.790{\times}10^{51}$ & $1.205(1.221)\,\keV$\\
\hline
 &   &   & $0.001$ & $6.630{\times}10^{45}\,\erg$ & $10.29\,\km$ & $1.288{\times}10^{55}$ & $321.3(961.1)\,\eV$\\
 \cline{4-8}
$1.4\,\msun$ & $2.032{\times}10^{57}$ & $10.31\,\km$ & $0.01$ & $6.592{\times}10^{48}\,\erg$ & $10.06\,\km$ & $1.278{\times}10^{56}$ & $32.19(96.14)\,\keV$\\\cline{4-8}
 &  &  & $0.1$ & $6.160{\times}10^{51}\,\erg$ & $7.535\,\km$ & $1.158{\times}10^{57}$ & $3.321(9.450)\,\MeV$\\
 \hline
\end{tabular}
\end{small}
\caption{\label{tab:1}\small Properties of \sqm stars.
Here, $\nb$ -- total baryon number, $R$ -- (areal) radius, $\ampl$ -- fractional amplitude of radial oscillations, $\erad$ -- energy to be eventually radiated away, $r_c$ -- corresponding conversion radius above which baryon density is lower than saturation one, $\meanexcen$ -- mean excitation energy per baryon, $\left.\excen\right|_R$ -- excitation energy per baryon at the surface.}
\end{table*} 	
	
One can calculate the change in mean baryon density for the \sqm object of constant density 
as  $\Delta n\eq n_0(1{-}(1{+}\ampl)^{\!{-}\!3})$. 
The \sqm planets of Earth-like masses are essentially such objects. 
In \tabref{tab:1} some properties of the Earth mass \sqm planet are listed. 
The radius  of this planet is $R\eq145.1\,\m$. 
For $0.1\%$ radial oscillations, $\ampl\eq0.001$, the surface displacement $\xi\eq x R\eq14.51\,\cm$, and $\Delta n\eq8.866{\times}10^{-4} \,\icfm$. 
The surface density decreases to $n_0{-}\Delta n$ and the baryon density inside the whole planet, $r{<}R$, decreases below the saturation density by almost the same amount. 
This is because the density gradient in the planet is very small and the central density differs little from the surface density $n_0$. 
Thus the whole planet becomes excited at the maximum expansion. 
The (mean) excitation energy is $E_\mathrm{exc}\eq1.205\,\keV$ per baryon. 
The baryon number of the planet $\nb\eq(4\pi{/}3)n_0 R^3\eq3.790{\times}10^{51}$. 
One can thus find that there is $7.315{\times}10^{42}\,\erg$ energy that could eventually be radiated away. 
The values listed in \tabref{tab:1} are found for a numerical model of an Earth mass \sqm. In the case of planets the energy can also be approximated with the help of analytical formula \eqref{eq:eradA} derived in the \appref.
		
For a real star one can estimate the change of the surface density by dividing the star into $N$ radial zones  of thickness $\Delta r\eq R/N$. 
The baryon number of each zone is conserved. 
The last zone has baryon number $\nb\eq4\pi\,n_0\, R^2\gammg(0) \Delta{r}$ (here, $\gammg$ is a relativistic correction factor). 
At the maximum expansion the radius changes to $R{+}\xi$ and $\nb\eq4\pi\,n(\xi)(R{+}\xi)^2(1{+}\xi/R)\gammg(\xi/R)\Delta{r}$. 
We thus find the baryon density at the surface to be $n(\xi)\eq\frac{n_0}{(1{+}\ampl)^3}{\cdot}\frac{\gammg(0)}{\gammg(\ampl)}$. 
For $R\eq10.31\,\km$, with the $10\%$ radius increase, $\ampl\eq0.1$ ($\xi\eq1.031\,\km$),  the baryon surface density is 	$0.2292\,\icfm$.

The amount of energy that can be released during one oscillation period is calculated within the MIT-bag model as follows. 
The value of the baryon density at the maximum expansion, $n(\xi)\eq0.2292\,\icfm$, corresponds to the energy per baryon $E\eq893.07\,\MeV$. 
At the saturation density the energy per baryon of the ground state of \sqm is $883.62\,\MeV$. 
The \sqm in the last shell thus exceeds the minimum energy by $9.450\,\MeV$ per baryon. 
To calculate all quark matter energy available for conversion into electromagnetic  energy, one should add all zones with baryon density less than ${\sim}0.39\,\icfm$ 
(reduced by a small and position dependent relativistic correction) in the equilibrium star as all these will have lower densities by a factor of ${\sim}0.75$, that is below saturation 
density $n_0$ at the maximum expansion. 
These zones comprise ${\sim}57\%$ of the star's baryon number and store the excitation energy of $6.160{\times}10^{51}\,\erg$. 
The excitation energy contained in the outermost shell of the star at the maximum expansion is $7.950{\times}10^{49}\erg$ for division of the star in 1000 zones. 
	
\section{THE FATE OF EXCITATION ENERGY}

Our main interest here is the fate of this excitation energy. The process of excitation of quark matter is an irreversible one. 
The excitation energy will be dissipated eventually into heat and radiation. 
Here we first focus on an intermediate step which is conversion of hadronic excitation energy into electromagnetic energy. 
To calculate  the generation rate of electromagnetic energy we need a time scale of  radial oscillations of the \sqm star. 
We  assume that the frequency of radial oscillations calculated in \mycitet{1990ApJ...363..603C} without considering any excitation of the \sqm matter can be used to obtain 
the first time scale of radial movement of the stellar matter just at the moment of the beginning of the first cycle of oscillations. 
We calculate the mean radial velocity of the surface, $v_s$, in the expansion phase. 
In a quarter of a period the surface moves outward a distance $\xi$, thus $v_s\eq\xi/(T/4)\eq xR/(T/4)\eq111\,\kmisec$ for $0.1\%$ ($\ampl\eq0.001$) oscillations. 
In terms of the radial coordinate of the stationary star in the same time the critical surface dividing excited and nonexcited matter moves inward reaching the radius of $r_c\eq10.29\,\km$ with velocity $v_c\eq(R{-}r_c){/}(T/4)\eq220\,\kmisec$. 
   	
The time scale for electromagnetic interactions is $10^{-16}\,\sec$ which is much shorter than $T/4\eq9.25{\times}10^{-5}\,\sec$ (the quarter of a period). 
Thus the excitation energy can be assumed to be converted into electromagnetic energy instantaneously. 
The rate of electromagnetic energy generation is $\epsrate\eq\erad/(T/4)$, where $\erad\eq6.630{\times}10^{45}\,\erg$ is the whole energy available for conversion, see \tabref{tab:1}. 
We find $\epsrate\eq7.17{\times}10^{49}\,\ergisec$. 
The mean excitation energy per baryon is $\meanexcen \eq\erad/\nb(r{>}r_c)$. 
Here $\nb(r{>}r_c)\eq1.288{\times}10^{55}$ is the number of baryons in the shell of excited \sqm, $r{>}r_c$. 
We find $\meanexcen \eq0.3213\,\keV$. 
Similarly, the numbers for $\ampl\eq0.1$, as shown in \tabref{tab:1}, are as follows: the total excitation energy is $\erad\eq6.160{\times}10^{51}\,\erg$, the critical surface radius is $r_c\eq7.535\,\km$, the number of excited baryons is $\nb(r{>}r_c)\eq1.158{\times}10^{57}$, the mean excitation energy per baryon is $\meanexcen \eq3.321\,\MeV$, the velocity is $v_c\eq3.004{\times}10^4\,\kmisec$, and finally the energy deposition rate is $\epsrate\eq6.66{\times}10^{55}\,\ergisec$. 
	
The above analysis concerned the process of energy conversion inside the \sqm star.
Astrophysically, it is interesting to know how much of the electromagnetic energy will be radiated away and how fast this will proceed. 
This problem requires further investigation. 
Here one may estimate conservatively the luminosity assuming that one half of the released energy in the outermost shell of thickness $\lambda$ is radiated away and the other half is absorbed by inner layers, where $\lambda$ is the photon mean free path.
	
The luminosity for $\ampl\eq0.001$ amplitude oscillations is $L\eq3.29{\times}10^{33}\,\ergisecifm{\cdot}\lambda$, where $\lambda$ must be expressed in $\fm$. 
For $\lambda\eq3.95\,\fm$,  $L\eq1.30{\times}10^{34}\,\ergisec$. 
The effective temperature defined through $L\eq4\pi R^2\sigma\Teff^4$ is $\Teff\eq2.0{\times}10^6\,\mathrm{K}$. 
The photon mean free path  $\lambda\eq3.95\,\fm$ corresponds to nondegenerate quark matter and reduced Thomson cross section $\sigma_T\eq(\frac{2}{3})^4(\frac{m_e}{m_u})^2 \cdot 665\,\mb$, where the up quark mass is $m_u c^2{\approx} 2\,\MeV$ and the electron mass $m_e c^2\eq0.511\,\MeV$. 
Thus $\sigma_T\eq8.58\,\mb$ and $\lambda\eq1/(\sigma_T n_0)$. 
Contributions from strange and down quarks are neglected here. 
When the quark matter is degenerate ($T\eq0$) the mean free path can be longer; however, it depends on the photon energy. 
The total energy released in the star, $\erad\eq6.6{\times}10^{45}\,\erg$ would sustain radiation with the luminosity $L\eq1.3{\times}10^{34}\,\ergisec$ for $\tau{\sim}\frac{\erad}{\lambda\,L}\eq1.6{\times}10^4$ years. 
However, neutrino cooling will switch on in $10^{-6}\,\sec$ and the star will cool fast. 
One can estimate order of magnitude of neutrino cooling time by using a standard value of neutrino emissivity of \sqm\ \mycitep{1990PhRvD..42..992G} $\epsnu\eq10^{24}\,\ergisecicc$. 
The  neutrino cooling time $\frac{\erad}{\epsnu\,\delta{V}}$ is $\tau_{\nu}\eq1.4{\times}10^{3}\,\sec$ ($\tau/\tau_{\nu}\eq3.5{\times}10^{8}$) for the volume of the star, or $\tau_{\nu}\eq1.5{\times}10^{5}\,\sec$ ($\tau/\tau_{\nu}\eq3.4{\times}10^{6}$) for the volume of the initial energy deposition shell. 
More accurate calculation requires taking into account the evolution of temperature of the \sqm star, which is beyond the scope of this research.

For $\ampl\eq0.1$ oscillations the surface luminosity is $L\eq1.5$ ${\times}10^{38}\,\ergisec$  and  time of radiation of $\erad\eq 6.16{\times}10^{51}\,\erg$ is $\tau\eq1.3{\times}10^6$ years. 
However, neutrino cooling time is $\tau_{\nu}\eq42$ years. 
As any excitation is damped in real stars, after some time the oscillations would cease to exist. 
In \mycitet{1990ApJ...363..603C} damping times for radial oscillations of compact stars are calculated. 
For the considered \sqm star of $1.4\,\msun$ the damping time is $5.47\,\sec$ \mycitep{1990ApJ...363..603C}. 
However, if the excitation mechanism considered here is accounted for, the whole energy of $10\%$ amplitude oscillations, of order of $10^{51}\,\erg$ is less than the value of excitation energy generated during the time of a single oscillation. 
Thus oscillations would be damped during the first cycle. 
Physically, this means that no radial oscillations of the \sqm star of such an amplitude would  exist in nature. 
For lower amplitudes the energy generation time is longer.
The damping effect on radial oscillations of \sqm stars by weak quark processes { {\it u}+{\it s} $\to$ {\it u}+{\it d} } are considered in 
\mycitet{WANG1984211} and \mycitet{PhysRevD.46.3290}. 
The process is very efficient, however, as a weak process it is not expected to determine the strong interaction phase transition we consider. 
But it could probably dominate the thermalization of excited \sqm.

The excitation energy of \sqm of density $n\eq (1\pls\ampl)^{-3}n_0$ is a form of latent heat. 
It will be released in the first-order phase transition with nucleation of the "new phase", which is the \sqm of saturation density $n_0$.
According to the simplest theory of first-order phase transition  the bubbles of saturation-density \sqm spontaneously form in the excited (lower-density-\sqm) phase. 
The minimal size of bubbles, $r_0$, is  {$r_0\eq{2\sigma}\dv\br{n\epsilon_{\mathrm{exc}}}$}, where $\sigma$ is the surface tension at the boundary of the two phases of \sqm. 
Once formed, the bubbles quickly grow  and enclose a higher and higher fraction of the surface. 
The latent heat is effectively released when the bubble walls collide with one another and then become thermalized.   
The surface tension $\sigma$, relevant to this phase transition is not available at present. 
However, on physical grounds, one can determine its limits, $0\lt\sigma\lt\sigma_0$, where $\sigma_0$ is the surface tension corresponding to the \sqm bag in an empty space. 
Its value in \citet{1984PhRvD..30.2379F} is constrained to be $60\lt\sigma_0^{1/3}\lt80\,\MeV$. 
   One can use a simple parameterization $\sigma\eq\sigma(n)\eq\sigma_0(1\mns n\dv n_0)$.  
The phase transition is governed by the strong interaction. 
It is thus a very fast process. 
The excitation energy (latent heat) is transferred to the quark matter during collisions of bubble walls and then thermalized by strong, electromagnetic, and weak interactions.

The amplitude of oscillations is physically determined by the excitation process in close encounters of \sqm objects and another compact star or black hole. 
This subject is studied thoroughly by \citet{2018PhRvD..98d4007Y}
in the case of quadrupole oscillations of neutron stars. 
Depending on the closest approach distance $R_c$ the energy transferred to the oscillations can reach $10^{53}\,\erg$  when $R_c{\sim}3R$, where $R$ is the star radius. 
By mode couplings, also monopole oscillations would be excited. 
Thus amplitudes of $10^{-6}$ are quite possible. 
The corresponding energy to be radiated away can be estimated from \eqref{eq:eradB}, which for a star of mass $1.4\,\msun$  predicts $\erad\eq6.63{\times}10^{36}\,\erg$ at $\ampl\eq10^{-6}$.
As far as higher multipole oscillations are concerned one should take into account the fact that quadrupole oscillations, by far the most important higher $l$ oscillations, tend to conserve the volume. 
It is thus quite unclear if these will {allow} density lower than $n_0$ {to occur}: one can imagine that the region{s} that expand are supplied with quark matter from shrinking regions{,} thus sustaining mean density of $n_0$ or higher.

\section{Concluding remarks}

There are many-fold astrophysical consequences of the instability of \sqm stars. 
The strongest tidal excitations are expected in very close neutron star binaries before their merger \mycitep{2010PhRvD..81b4012B}. 
Such objects are prime candidates for gravitational wave (GW) sources. 
If one of the stars is the \sqm star then the radial oscillations excited by the companion star could destroy the \sqm star before the merger and the GW emitted during last orbits would differ from the templates. 
The neutrino cooling may not be fast enough to prevent the fragmentation of the excited \sqm matter into smaller \sqm objects. 
Fragmentation would proceed down to fragments that can cool effectively. 
Distribution of fragments in \sqm objects decay is calculated in \mycitet{2014PhLB..733..164P}. 
If both stars are \sqm stars, then they mutually excite one another and thus both can be destroyed before the merger. 
The fragmentation of \sqm stars in close binaries was studied by 
	\cite{PhysRevD.46.3290}. 
		
For a $1\,\mearth$ \sqm planet of radius $R\eq145\,\m$ radial oscillation of $10\%$ amplitude, $\xi\eq1.45\,\m$,  give baryon density $0.7513\,n_0$ and excitation energy of $11.84\,\MeV$. 
Total excitation energy is $\erad\eq7.187{\times}10^{46}\,\erg$. 
This makes $1.340\%$  of the rest energy. 
Also radial oscillations of \sqm planets excited in binary systems could lead to disastrous consequences if they are of the same relative amplitude as that for massive \sqm stars. 

We studied homogeneous \sqm stars without crusts. But
the \sqm stars can be surrounded by a crust of usual neutron star matter of rather small mass and of densities below neutron drip density 
\citep{Huang1997a, Huang1997b}.
This layer, as a solid, would suffer cracks and disruption at higher amplitudes of oscillations. 
More importantly, the crust would absorb electromagnetic energy radiated by the \sqm core. 
It would heat up and become a source of thermal radiation from its surface. 
However, a further study is needed to accurately account for its influence {on} the el-mag spectra of the star.

In conclusion, \sqm stars and planets are very sensitive to radial oscillations. 
Even $0.1\%$  oscillations of the radius result in about $1\,\keV$ per unit baryon number excitation energy in the surface layer of every \sqm object, equally for stars of pulsar masses and planetary-mass \sqm objects. 
In the evolution of binary systems with the \sqm objects the energy loss due to excitations of  \sqm stars and planets must be accounted for and this can change significantly the predictions obtained with unexcited \sqm objects. 
This is particularly relevant to binary gravitational wave sources  as the exact template of the system gravitational amplitude cannot be calculated without including excitations of \sqm objects. 
	
\appendix

\section{\label{app:}Scaling laws for energy excited by monopole perturbation }
SQM matter can be characterized at saturation point by three numbers: the saturation baryon density $n_s$, the minimum energy per baryon $\epsat$ and convexity parameter $\beta_s$. 
As determined from the \sqm equation of state:
$$n_s\eq0.296136\,\icfm,\quad\epsat\eq883.623\,\MeV,\quad \beta_s\!\equiv\!\frac{n_s^2 \epsilon''(n_s)}{\epsat}\eq0.307124.$$
The excess energy per baryon $\Delta \epsilon$ above $\epsat$  due to baryon number density variation $\Delta n$ about the saturation point $n\eq n_s$ is $\Delta \epsilon \eq\frac{1}{2}\epsilon''(n_s)(\Delta n)^2$, because $\epsilon'(n_s)\eq0$ by definition of that point. 

For small mass stars (planets), $n(r)$ is almost constant and equal to baryon density at the surface, $n(r)\!\sim\! n_s$. 
Upon scaling $r\!\to\!r\xiplus$ with ${0\lt\ampl\!\ll1}$, the density decreases to $n_s/\xiplus^3$ by the amount $\Delta n\eq 3 n_s \ampl$, resulting in energy per baryon increase $\Delta \epsilon\eq\frac{9}{2} \beta \epsat \ampl^2\eq1221.22\,\ampl^2\,\MeV$ in whole volume of the star. 
On
multiplying $\Delta\epsilon$ by total number of {baryons} $\nb\eq  Mc^2/\epsat\eq 3.794{\times} 10^{51} M/\mearth$ one obtains
an estimation of total energy to be radiated away 
\begin{equation}\label{eq:eradA}\erad\eq\frac{9}{2}\beta_s M c^2 \ampl^2\eq 7.416 {\times} 10^{48}\,\erg\cdot\frac{M}{\mearth}\,\ampl^2.\end{equation}
The energy scales as $\ampl^2$. 
In particular, $\erad\eq 7.4{\times}10^{38}\,\erg$ for an Earth mass \sqm planet undergoing radial deformation of amplitude $\ampl\eq10^{-5}$.

\begin{figure}[h]
\centering
\begin{tabular}{@{}ll@{}}
$\phantom{x}$A) & $\phantom{x}$B)\\
\includegraphics[trim=0 0 0 0,clip,width=0.49\columnwidth]{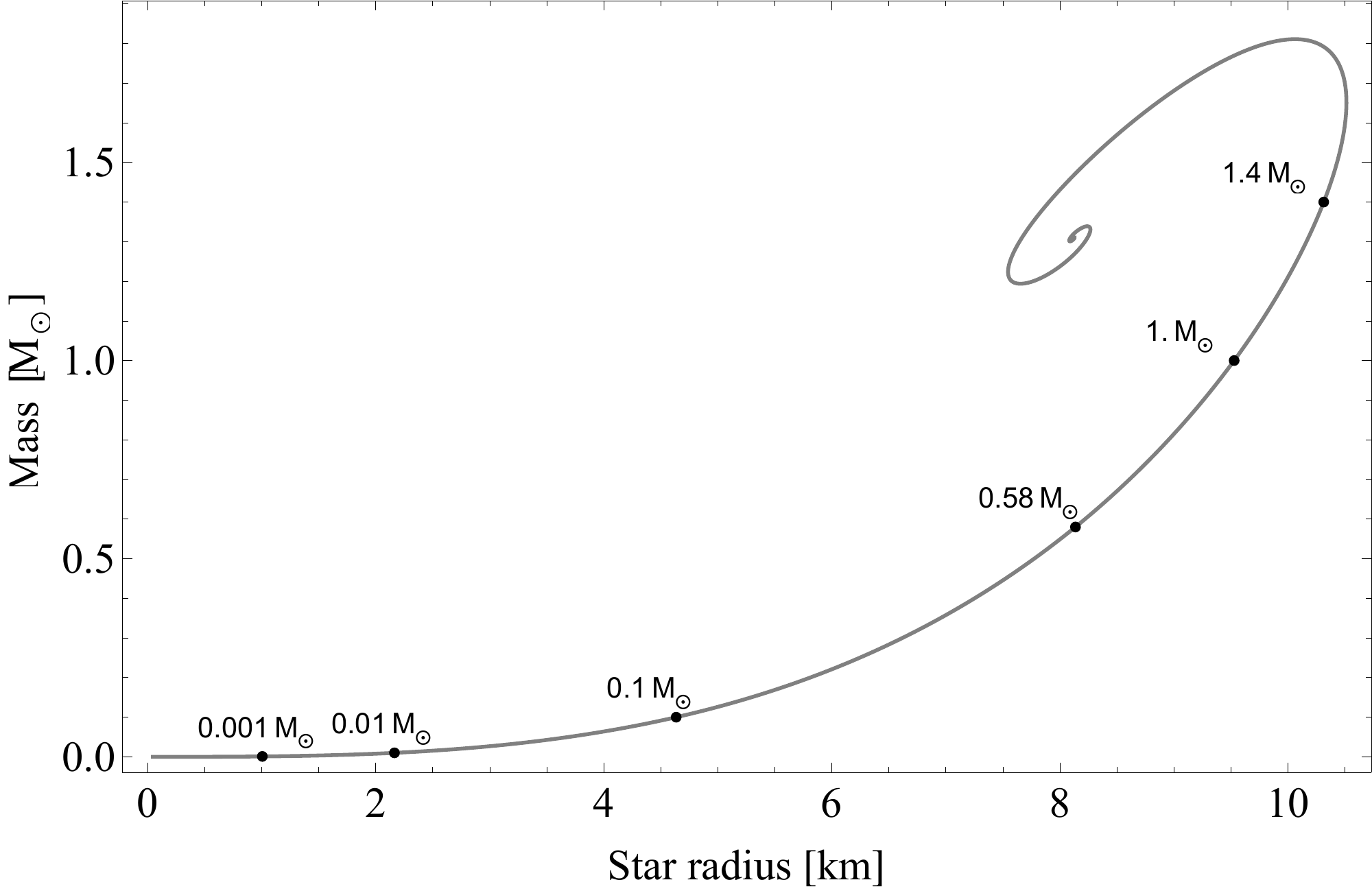}&
\includegraphics[trim=0 0 0 0,clip,width=0.49\columnwidth]{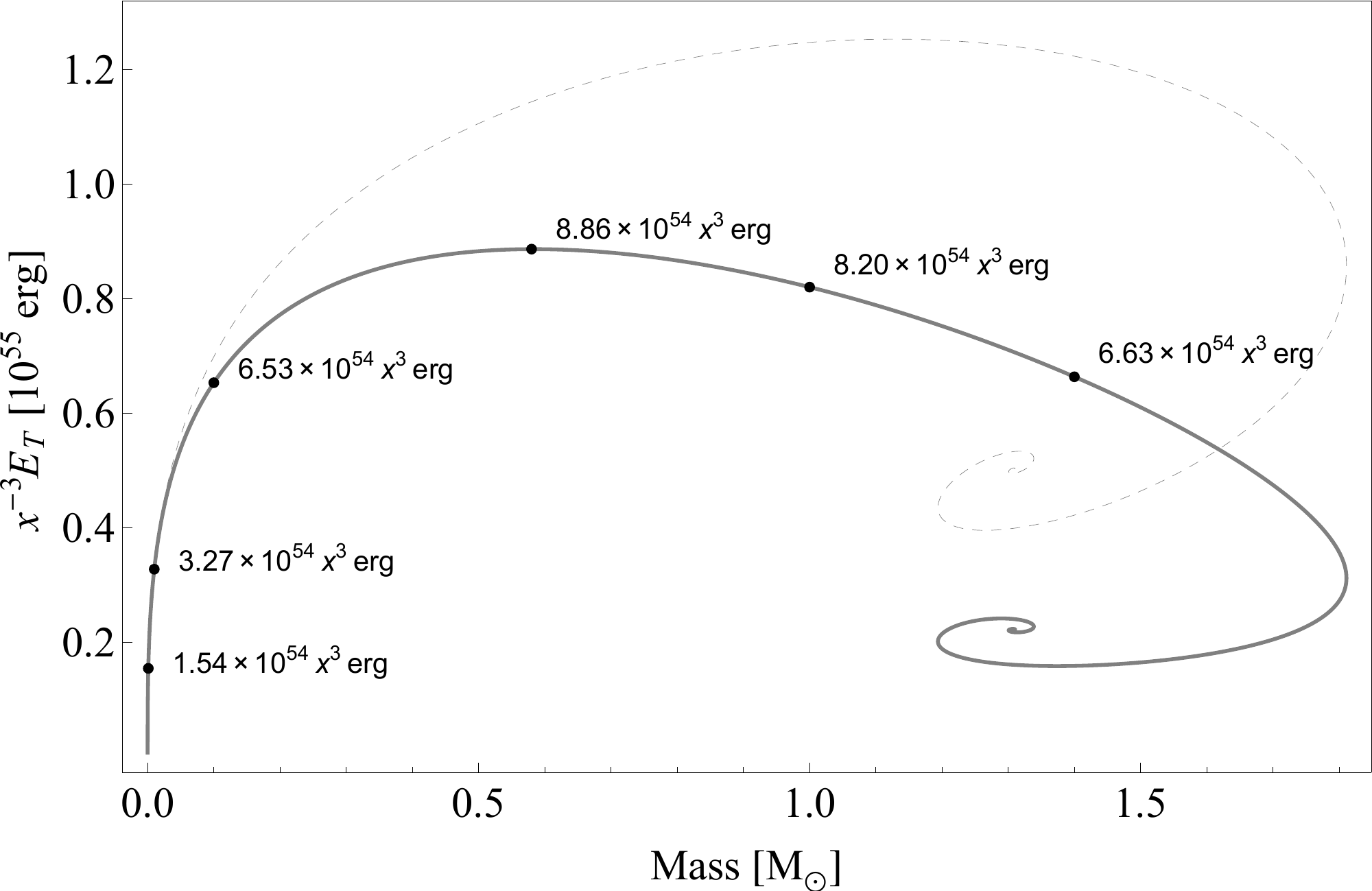}%
\end{tabular}
\caption{\label{fig:rmerad}\small {\it Left panel:} Mass-radius relation for \sqm objects ($B\eq60\,\mevicfm$, $\ms\eq150\,\MeV$) as determined from the numerical steady-state solution of a relativistic spherical star. 
{\it Right panel:} Energy defined in \eqref{eq:eradB} to be released by \sqm star  excited in a thin layer at the star surface by monopole mode of amplitude $\ampl$ ($B\eq60\,\mevicfm$, $\ms\eq150\,\MeV$). 
For comparison, the thin dashed line disregards the relativistic correction $H_{\Phi}$. 
}
\end{figure}

The numerical model shows that by keeping $\ampl$ constant, the subsaturation zone shrinks toward the surface as the mass increases. 
In the limiting case, only baryons in the thin zone of some width $\xiplus\,\dR$ 
should be considered. 
In this zone, the scaled density profile $n_{\ampl}(h)\!\approx\! n_s\br{1{-}(3{+}\Gamma^2\Phi)\ampl}\!-\!h\,n'(R)\dR$ (as obtained in first-order expansion in $\ampl$ and $\dR$) is used to approximate the distribution of baryons in the deformed star surface vicinity, where $\Phi\!\equiv\!\frac{G M}{Rc^2}$, $\Gamma\!\eq\! \br{1{-}2\Phi}^{\!-\!1\!{/}\!\!\;2}$ and $h$ parameterizes the position (with $h\eq0$  representing the outer and  $h\eq1$ the inner boundary of that zone). 
For the condition $n_{\ampl}(1)\eq n_s$ to be satisfied at the surface of the expanded star,
one needs $\dR\!\eq{-}\!\br{3\!+\!\Gamma^2\Phi}\!\ampl\,n_s/n'(R)$ to the leading order in $\ampl$. 
Making use of the general relativistic equation for static perfect fluid spheres allows us to determine $n'(R)$ at the surface of the undisturbed star. 
Considering that $\rho(n)\eq n\,\epsilon(n)$, $n(R)\eq n_s$ and $p(n_s)\eq 0$, 
the condition reduces to $R\,n'(R)p'(n_s)\eq{-}n_s\epsat \Gamma^2\Phi$. 
On differentiating  $\rho'(n)\eq (\rho+p)/n$ with respect to $r$ one obtains $p'(n_s)\eq\beta_s \epsat$, 
hence $R\,n'(R)\eq{-}\frac{n_s}{\beta_s}\Gamma^2\Phi$, 
and finally $\dR/R\eq3\beta_s\frac{c^2R}{G M}\!\br{\!1{-}\frac{7}{3}\frac{GM}{Rc^2}}\!\ampl$, hence  $\Delta E_\mathrm{rad}\eq 4\pi\!\int_{\xiplus(R-\dR)}^{\xiplus R} \Gamma(\tilde{r})  \tilde{r}^2n_\ampl(\tilde{r})\br{\epsilon(n_\ampl(\tilde{r})){-}\epsat}\ud{\tilde{r}}$. 
Changing variables $\tilde{r}\eq(1+\ampl)r$, expanding the integrand about $r\eq R$ and noticing that $r{-}R\eq\mathcal{O}(\ampl)$, we obtain  to the leading order in $\ampl$:
\begin{equation}\label{eq:eradB}
 \erad\eq18\pi\beta_s^2\epsat n_s \frac{c^2R^4}{GM}{\cdot}H_{\Phi}{\cdot}\ampl^3\quad \mathrm{if}\quad \frac{\dR}{R}\equiv\frac{3\beta_s}{\Phi}\!\br{\!1{-}\frac{7}{3}\Phi}\!\ampl\ll1, \end{equation} 
 where   $H_{\Phi}\eq\sqrt{1{-}2\Phi}\left(\!\frac{1{-}\frac{7}{3}\Phi}{1{-}2\Phi}\!\right)^{\!\!3}$ is a relativistic correction factor unimportant for small mass stars. The energy scales as $\ampl^3$.
  For $M\eq1.4\msun$, $R\eq10.31\,\km$, $\dR\dv R\eq2.75\,\ampl$ and $\erad\eq6.634{\times}10^{54}\ampl^3\,\erg$. 
  The function $\erad(M(R))$ for a whole branch of \sqm stars can be seen in \figref{fig:rmerad}.
  
\acknowledgements

\noindent
{\bf Acknowledgements.} We gratefully acknowledge the referee for constructive advices and suggestions which
have contributed to improve this paper.

\bibliography{referencjechg}

\end{document}